\documentclass[conference]{IEEEtran}
\IEEEoverridecommandlockouts
\usepackage{cite}
\usepackage{amsmath,amssymb,amsfonts}
\usepackage{algorithmic}
\usepackage{graphicx}
\usepackage{textcomp}
\usepackage{xcolor}
\usepackage{multirow}
\def\BibTeX{{\rm B\kern-.05em{\sc i\kern-.025em b}\kern-.08em
    T\kern-.1667em\lower.7ex\hbox{E}\kern-.125emX}}
\begin{document}
\title{HMT-UNet: A hybird Mamba-Transformer Vision UNet for Medical Image Segmentation
}

\author{
Mingya Zhang$^{1} $  \quad Liang Wang$^{1} $ \quad Zhihao Chen$^{2}$ \quad Yiyuan Ge$^{2}$ \quad Xianping Tao$^{1}$\\
    $^{1}$ Nanjing University \quad
    $^{2}$ BISTU \quad
}

\maketitle

\begin{abstract}
In the field of medical image segmentation, models based on both CNN and Transformer have been thoroughly investigated. However, CNNs have limited modeling capabilities for long-range dependencies, making it challenging to exploit the semantic information within images fully. On the other hand, the quadratic computational complexity poses a challenge for Transformers. State Space Models (SSMs), such as Mamba, have been recognized as a promising method. They not only demonstrate superior performance in modeling long-range interactions, but also preserve a linear computational complexity. 
The hybrid mechanism of SSM (State Space Model) and Transformer, after meticulous design, can enhance its capability for efficient modeling of visual features. Extensive experiments have demonstrated that integrating the self-attention mechanism into the hybrid part behind the layers of Mamba's architecture can greatly improve the modeling capacity to capture long-range spatial dependencies.
In this paper, leveraging the hybrid mechanism of SSM, we propose a U-shape architecture model for medical image segmentation, named Hybird Transformer vision Mamba UNet (HTM-UNet).
We conduct comprehensive experiments on the ISIC17, ISIC18, CVC-300, CVC-ClinicDB, Kvasir, CVC-ColonDB, ETIS-Larib PolypDB public datasets and ZD-LCI-GIM private dataset.
The results indicate that HTM-UNet exhibits competitive performance in medical image segmentation tasks.
Our code is available at https://github.com/simzhangbest/HMT-Unet.
\end{abstract}


\section{Introduction}
\label{sec:intro}

\begin{figure*}
    \centering
    \includegraphics[width=0.9\textwidth]{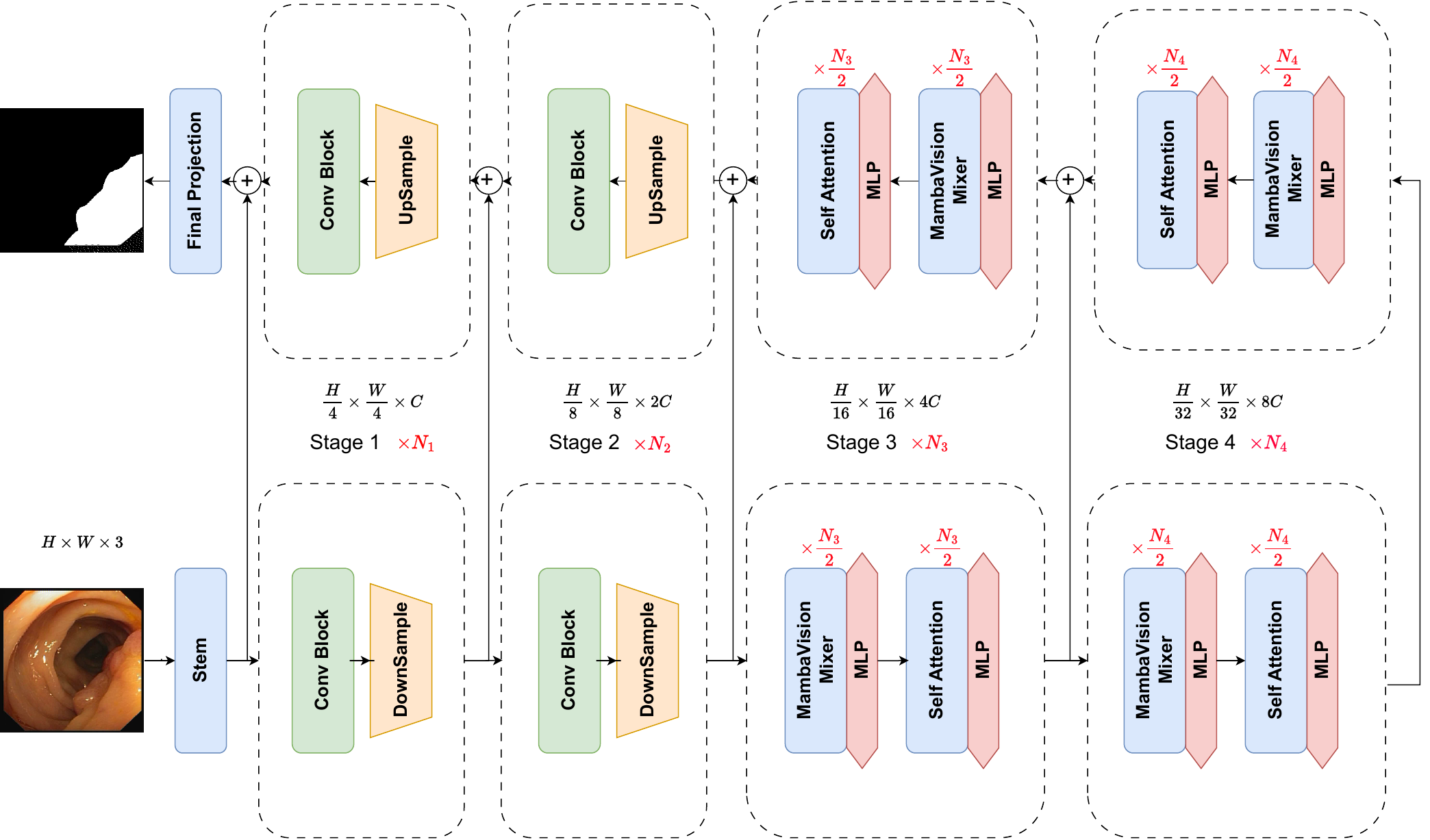}
    \caption{HMT-Unet can be divided into three parts: the encoder, the decoder, and the skip connection. The encoder consists of convolutional modules and MambaVision Mixer , The decoder is composed of MambaVision Mixer, upsampling operations, convolutional modules, and a final linear layer. }
    \label{fig:01}
\end{figure*}

As medical imaging technology continues to advance, medical images have become a crucial tool for diagnosing diseases and planning treatments~\cite{cheng2016computer}. Among the fundamental and critical techniques in medical image analysis, medical image segmentation holds a significant place. This process involves distinguishing pixels of organs or lesions in medical images, such as CT scans~\cite{golan2016lung} and Endoscopy~\cite{tang2020development} videos. 
Medical image segmentation is one of the most difficult tasks in medical image analysis, with the goal of providing and extracting vital information regarding the shape of these organs or tissues.
% Deep learning techniques have been used recently to improve medical image segmentation. These models extract useful information from images, increase accuracy, and adapt to different datasets and tasks.

A common approach for semantic image segmentation is the use of an Encoder-Decoder network with skip connections. In this framework, the Encoder captures hierarchical and abstract features from an input image. The Decoder, on the other hand, uses the feature maps produced by the Encoder to build pixel-wise segmentation masks, attributing a class label to each pixel in the input image. Numerous studies have been done to integrate global information into feature maps and enhance multi-scale features, leading to significant enhancements in segmentation performance~\cite{chen2021transunet,zhou2018unet++,liu2018path}.

Transformers~\cite{vaswani2017attention} have revolutionized various fields such as computer vision, natural language processing, speech, and robotics, largely due to their adaptive attention mechanism and multimodal adaptability. However, their computational demand grows quadratically with sequence length, posing challenges for training and application.
The innovative State Space Model(SSM) introduced by Mamba~\cite{gu2023mamba} offers a solution with its linear time complexity, rivaling or surpassing Transformer performance in language tasks. Mamba's key innovation lies in its efficient, hardware-conscious selection mechanism for processing lengthy sequences, addressing the scalability issues of traditional Transformers.

Vision Mamba (Vim)~\cite{zhu2024vision} employs bidirectional SSMs to overcome Transformer deficiencies in capturing global context and spatial relationships. However, the latency introduced by processing sequences for bidirectional SSMs can complicate training, risk overfitting, and doesn't guarantee enhanced accuracy. Despite these challenges, ViTs and CNNs often surpass Mamba-based models in visual tasks due to their efficiency and reliability.

Recently, models that integrate Mamba and Transformer have garnered significant interest among researchers. NVIDIA has introduced MambaVision~\cite{hatamizadeh2024mambavision}, an innovative hybrid model specifically designed for computer vision tasks, combining the strengths of Mamba and Transformer architectures. It features a hierarchical structure with multi-resolution design, utilizing CNN-based residual blocks to rapidly extract features at various resolutions.
Inspired by the success of MambaVision in image classification tasks, this paper introduces a MambaVision-based UNet model for medical segmentation for the first time, the \textbf{H}ybrid \textbf{M}amba-\textbf{T}ransformer Vision UNet (HMT-Unet).
The main contributions of this paper can be summarized as follows:
% We carry out exhaustive experiments on tasks related to gastroenterology semantic segmentation task and skin lesion segmentation to showcase the capabilities of HMT-Unet in the field of medical image segmentation. 

\textbf{1.} We propose HMT-UNet, marking the first occasion of exploring the potential applications of purely hybird models of SSM and Transformer in medical image segmentation.

\textbf{2.} Comprehensive experiments are conducted on public and private datasets, with results indicating that HMT-UNet exhibits considerable competitiveness.

\textbf{3.} We establish a baseline for Hybird of Mamba and  Transformer (HMT) models in medical image segmentation tasks, providing valuable insights that pave the way for the development of more efficient and effective SSM-based segmentation methods.

\section{Methods}

\begin{figure}
    \centering
    \includegraphics[width=0.6\linewidth]{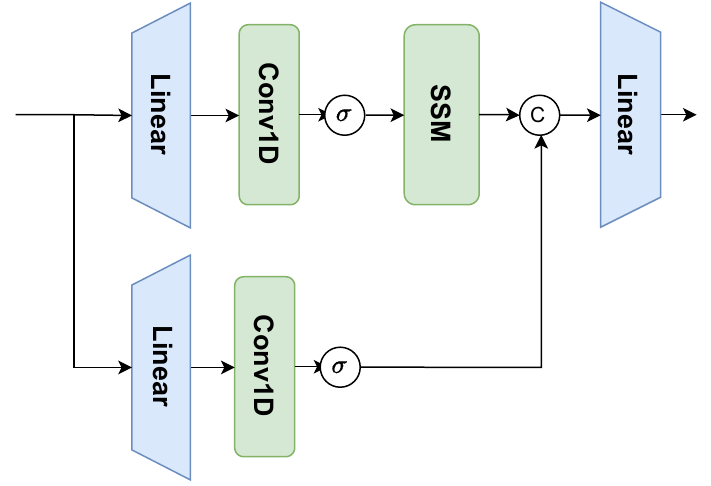}
    \caption{The architecture of MambaVision block. A symmetric path without SSM as a token mixer to enhance the modeling of global context.}
    \label{fig:2}
\end{figure}

\subsection{HMT-Unet Overall Architecture}
As shown in Fig~\ref{fig:01}. HMT-Unet can be divided into three parts: the encoder, the decoder, and the skip connection. The encoder consists of convolutional modules from MambaVision and MambaVision Mixer (a hybrid module of VSSM and Transformer), which perform downsampling as the network deepens. The decoder is composed of MambaVision Mixer, upsampling operations, convolutional modules, and a final linear layer that serves to restore the dimensions of the segmentation results. For the skip connection, to showcase the segmentation performance of the purest SSM-based model, we have employed only the simplest addition operation.

The encoder and decoder are structurally symmetric, with the first two layers being CNN-based layers for fast feature extraction at higher input resolutions, and the subsequent two layers being blocks that combine MambaVision and Transformer.
Specifically, given an image of size $H \times W \times 3$, the input is first converted into overlapping patches with size $\frac{H}{4} \times \frac{W}{4} \times C$ and projected into a $C$ dimensional embedding space by the stem which consists of two consecutive $3 \times 3$ CNN layers with stride of $2$. The downsampler in between stages consists of a batch normalized $3 \times 3$ CNN layer with stride $2$ which reduces the image resolution by half. Stage $3$ and $4$ employ both MambaVision and Transformer blocks. Specifically, given $N$ layers, we use $\frac{N}{2}$ MambaVision and MLP blocks which are followed by additional $\frac{N}{2}$ Transformer and MLP blocks.

In the decoder, each layer in stage $4$ and stage  $3$ must first go through the MambaVision Mixer before being passed to the Transform module. the convolution in stages $2$ is also $3 \times 3$ CNN layer with stride $2$ and sample layers, and the Upsample section uses linear upsampling to double the feature size.

\subsection{Mamba Preliminaries}
In contemporary SSM-based models, namely, Structured State Space Sequence Models (S4) and Mamba~\cite{gu2023mamba,liu2024vmamba,ruan2024vm}, both depend on a traditional continuous system that maps a one-dimensional input function or sequence, represented as $x\left ( t \right ) \in  R $, through intermediary implicit states $h\left ( t \right ) \in  R^{N}$ to an output $y\left ( t \right ) \in  R $. This process can be depicted as a linear Ordinary Differential Equation (ODE):

\begin{equation}
\begin{split}
         &h{}'\left ( t \right )  =Ah\left ( t \right ) +Bx\left ( t \right ) \\
         &y\left ( t \right ) =Ch\left ( t \right )
\end{split}
\end{equation}
where $\mathbf{A} \in R^{N \times N} $ represents the state matrix, while $\mathbf{B} \in R^{N \times 1} $ and $\mathbf{C} \in R^{N \times 1} $ denote the projection parameters.

The timescale parameter $\boldsymbol{\Delta}$ and convert $\mathbf{A}$ and $\mathbf{B}$ into discrete parameters  $\mathbf{\overline{A}}$ and $\mathbf{\overline{B}}$ using a consistent discretization rule. The zero-order hold (ZOH) is typically utilized as the discretization rule and can be outlined as follows:
\begin{equation}
\begin{split}
        &\overline
{\mathbf{A}}=\exp (\boldsymbol{\Delta} \mathbf{A}) \\
        &\overline{\mathbf{B}}=(\boldsymbol{\Delta} \mathbf{A})^{-1}(\exp (\boldsymbol{\Delta} \mathbf{A})-\mathbf{I}) \cdot \boldsymbol{\Delta} \mathbf{B}
\end{split}
\end{equation}

Following discretization, SSM-based models can be calculated in two distinct methods: linear recurrence or global convolution, which are denoted as equations \eqref{eq:1} and \eqref{eq:2}
\begin{equation}
    \begin{split}
        &h{}' \left ( t \right ) = \overline{\mathbf{A}}  h\left ( t \right ) + \overline{\mathbf{B}}x\left ( t \right ) \\
        &y\left ( t \right ) =\mathbf{C}h\left ( t \right )  \label{eq:1}
    \end{split}
\end{equation}
\begin{equation}
    \begin{split}
        & \overline{K} = \left ( \mathbf{C}\overline{\mathbf{B}} , \mathbf{C}\overline{\mathbf{A}\mathbf{B}} ,...,\mathbf{C}\overline{\mathbf{A} } ^{L-1}\overline{\mathbf{B}}    \right )  \\
        & y=x*\overline{\mathbf{K} } \label{eq:2}
    \end{split}
\end{equation}
where $\overline{\mathbf{K} } \in R^{L} $ represents a structured convolutional kernel, and $L$ denotes the length of the input sequence $x$.

\subsection{MambaVision Layer and Mixer}
Assuming an input $X \in \mathbb{R}^{T \times C}$ with sequence length $T$ with embedding dimension $C$, the output of layer $n$ in stages 3 and 4 can be computed as in
\begin{equation}
\begin{array}{l}
\hat{{X}}^{n}=\text{Mixer}(\text{Norm}({X}^{n-1}))+{X}^{n-1} \\
{X}^{n}=\text{MLP}(\text{Norm}(\hat{{X}}^{n}))+\hat{{X}}^{n}, \\
\end{array}
\label{eq:eq_layer}
\end{equation}
Norm and Mixer denote the choices of layer normalization and token mixing blocks, respectively. Without loss of generality, Layer Normalization is used for Norm. Given $N$ layers, the first $\frac{N}{2}$ layers employ MambaVision mixer blocks while the remaining $\frac{N}{2}$ layers employ self-attention. 

As shown in Fig.~\ref{fig:2}, we directly employ the MambaMixer proposed by MambaVision~\cite{hatamizadeh2024mambavision} to perform the Mamba Vision SSM operations. It is worth noting that the causal convolution in this module has been replaced with regular convolution, as the characteristic of causal convolution, which limits the influence to one direction, is not important in vision tasks. An additional symmetric branch without SSM has been added to the Mixer module, consisting of an additional convolution and SiLU activation, to compensate for any content lost due to the sequential constraints of SSMs. Then, these two branches are concatenated and fed into the final linear layer.

This combination ensures that the final feature representation incorporates both the sequential and spatial information, leveraging the strengths of both branches. We note that the output of each branch is projected into an embedding space with size $\frac{C}{2}$ (\textit{i.e.} half the size of original embedding dimension) to maintain similar number of parameters to the original block design. Given an input $X_{in}$, the output of MambaVision mixer $X_{out}$is computed according to
\begin{equation}
\begin{array}{l}
{X_{1}}=\text{Scan}(\sigma(\text{Conv}(\text{Linear}(C, \frac{C}{2})({X_{in}})))) \\
{X_{2}}=\sigma(\text{Conv}(\text{Linear}(C, \frac{C}{2})({X_{in}}))) \\
{X_{out}}=\text{Linear}(\frac{C}{2},C)(\text{Concat}(X_{1}, X_{2})),
\end{array}
\label{eq:eq_mixer1}
\end{equation}
$\text{Linear}(C_{in}, C_{out})(\cdot)$ denotes a linear layer with $C_{in}$ and $C_{out}$ as input and output embedding dimensions, $\text{Scan}$ is the selective scan operation as in \cite{gu2023mamba} and $\sigma$ is the activation function for which Sigmoid Linear Unit (SiLU)~\cite{elfwing2018sigmoid} is used. In addition, \text{Conv} and \text{Concat} represent 1D convolution and concatenation operations.  

\subsection{Self-attention}
We use a generic multihead self-attention mechanism in accordance to
\begin{equation}
    {\rm Attention}({Q}, {K}, {V}) = {\rm Softmax}(\frac{{Q}{K}^\mathsf{T}}{\sqrt{d_{h}}}){V}.
    \label{eqn:mhsa}
\end{equation}
$Q, K, V$ denote query, key and value respectively and $d_{h}$ is the number of attention heads. Without loss of generality, the attention formulation which can be computed in a windowed manner similar to previous efforts~\cite{liu2021swin, liu2022swin}.

\subsection{Loss function}

For our medical image segmentation tasks, we primarily employ basic Cross-Entropy and Dice loss as the loss function cause all of our dataset masks comprise two classes, which are a singular target and the background.

\begin{equation}
    \begin{split}
        &L_{\text {BceDice }}=\lambda_{1} L_{\mathrm{Bce}}+\lambda_{2} L_{\text {Dice }} \\
        &L_{Bce} = -\frac{1}{N}\sum_{1}^{N}\left [ y_{i} log\left (  \hat{y}_{i} \right ) +\left (  1-y_{i}\right )log\left ( 1-\hat{y}_{i}  \right )   \right ] \\
        &L_{\text {Dice }}=1-\frac{2|X \cap Y|}{|X|+|Y|}
    \end{split}
\end{equation}
$\left ( \lambda_{1}, \lambda_{2} \right ) $ are constants, with $\left ( 1, 1 \right ) $ often selected as the default parameters.

\section{Experiments}
\label{experiments}

\subsection{Datasets}

% \begin{table}[ht]
% \caption{In the field of gastroenterology, commonly used publicly available datasets for medical image segmentation are listed in the table, along with their image quantities and sizes.}
% \label{tab:datasets}
% \centering

% \begin{tabular}{lll}
% \hline
% \textbf{Dataset} & \textbf{Numbers} & \textbf{Size} \\ \hline
% Kvasir-SEG & 1000 & Variable \\
% CVC-ClinicDB & 612 & $384\times288$ \\
% Endoscene & 912 & $574\times500$ \\
% CVC-ColonDB & 380 & $574\times500$ \\
% ETIS & 196 & $1255\times966$ \\
% ZD-LCI-GIM & 1020 & $1280\times1024$ \\ \hline
% \end{tabular}%

% \end{table}

\textbf{Skin lesion datasets:} The International Skin Imaging Collaboration 2017 and 2018 challenge datasets (ISIC17 and ISIC18) \cite{isic17,isic18} are two publicly available skin lesion segmentation datasets, containing 2,150 and 2,694 dermoscopy images with segmentation mask labels, respectively.

\textbf{Gastrointestinal polyp datasets:} The Kvasir-SEG~\cite{jha2020kvasir}, ClinicDB~\cite{bernal2015wm}, ColonDB~\cite{tajbakhsh2015automated}, Endoscene~\cite{vazquez2017benchmark}, and ETIS~\cite{silva2014toward} are currently publicly available polyp datasets.
 % with the number of images and image sizes corresponding to each dataset as shown in Table~\ref{tab:datasets}.

% 第三类是我们的私有数据集，胃部肠化数据集 ZD-LCI-GIM, 是用来观察胃部早期癌变可能性的重要的胃肠镜图片数据

\textbf{Gastrointestinal GIM dataset:} 
% As shown in Table~\ref{tab:datasets}, 
The ZD-LCI-GIM dataset of gastric intestinal metaplasia, which is an important gastroscopic image data set for observing the possibility of early gastric cancer. This part of the dataset has been proposed by us and will be open-sourced soon.

For these datasets, we provide detailed evaluations on several metrics, including Mean Intersection over Union(IOU), Dice Similarity Coefficient(DSC), Accuracy(Acc), Sensitivity(Sen), and Specificity(Spe).

\subsection{Experimental Setup}
We adjust the image dimensions in all datasets to $256 \times 256$ pixels, We also bring in data augmentation methods, such as random flipping and random rotation. 
In terms of operational parameters, we have the batch size set at 80, with the AdamW optimizer engaged starting with a learning rate of 1e-3. We make use of CosineAnnealingLR as the scheduler, with its operation spanning a maximum of 50 iterations and the learning rate going as low as 1e-5. We conduct our training over the course of 200 epochs.
For the HMT-Unet, the backbone encoder units' weights are initially set to align with those of MambaVsion-S. The implementation was carried out on an Ubuntu 20.04 system, using Python 3.9.12, PyTorch 2.0.1, and CUDA 11.7, All experiments are conducted on a single NVIDIA RTX V100 GPU.

\subsection{Results}

We compare HMT-Unet with some state-of-the-art models, presenting the experimental results in Table~\ref{tab:isic_all}, Table~\ref{tab:kc}, Table~\ref{tab:CEE} and Table~\ref{tab:gim}.
For the ISIC and ZD-LCI-GIM datasets, our VM-UNetV2 outperforms other models in terms of the IOU, DSC and Acc metrics. 
% 我们的模型在息肉数据集上的表现
On the Kvasir-SEG, ClinicDB, and ETIS datasets, our algorithm achieved state-of-the-art (SOTA) performance, and it also showed competitive performance on ColonDB and Endoscene.

% 实验数据存放地址：/root/workspace/code/mamba_all/VM-UNet/results/hmt_unet_polyp_Monday_19_August_2024_21_57_28						
% /root/workspace/code/mamba_all/VM-UNet/results/hmt_unet_polyp_Monday_19_August_2024_13_36_31
% /root/workspace/code/mamba_all/VM-UNet/results/hmt_unet_isic_all_Monday_19_August_2024_21_55_05
\begin{table}[hbt]
\caption{Comparative experimental results on the ISIC17 and ISIC18 datasets(Bold indicates the best)}
\label{tab:isic_all}
\resizebox{\columnwidth}{!}{%
\begin{tabular}{cccccc}
\hline
\textbf{Model}     & \textbf{IOU} & \textbf{DSC} & \textbf{Acc} & \textbf{Spe} & \textbf{Sen} \\ \hline
UNet~\cite{ronneberger2015u}               & 76.98              & 86.99             & 95.65             & 97.43             & 86.82             \\
UTNetV2~\cite{gao2022multi}            & 77.35              & 87.23             & 95.84             & 98.05             & 84.85             \\
TransFuse~\cite{zhang2021transfuse}          & 79.21              & 88.40             & 96.17             & 97.98             & 87.14             \\
MALUNet~\cite{ruan2022malunet}            & 78.78              & 88.13             & 96.18             & 98.47             & 84.78             \\
UNetV2~\cite{peng2023u}             & 82.18              & 90.22             & 96.78             & 98.40             & 88.71             \\
VM-UNet~\cite{ruan2024vm}            & 80.23              & 89.03             & 96.29             & 97.58             & 89.90             \\
\textbf{HMT-Unet} & \textbf{83.05}     & \textbf{90.74}    & \textbf{96.88}    & 98.02             & \textbf{91.21}    \\ \hline
UNet~\cite{ronneberger2015u}               & 77.86              & 87.55             & 94.05             & 96.69             & 85.86             \\
UNet++~\cite{zhou2018unet++}             & 78.31              & 87.83             & 94.02             & 95.75             & 88.65             \\
Att-Unet~\cite{oktay2018attention}           & 78.43              & 87.91             & 94.13             & 96.23             & 87.60             \\
UTNetV2~\cite{gao2022multi}            & 78.97              & 88.25             & 94.32             & 96.48             & 87.60             \\
SANet~\cite{wei2021shallow}              & 79.52              & 88.59             & 94.39             & 95.97             & 89.46             \\
TransFuse~\cite{zhang2021transfuse}          & 80.63              & 89.27             & 94.66             & 95.74             & 91.28             \\
MALUNet~\cite{ruan2022malunet}            & 80.25              & 89.04             & 94.62             & 96.19             & 89.74             \\
UNetV2~\cite{peng2023u}             & 80.71              & 89.32             & 94.86             & 96.94             & 88.34             \\
VM-UNet~\cite{ruan2024vm}            & 81.35              & 89.71             & 94.91             & 96.13             & 91.12             \\
\textbf{HMT-Unet} & \textbf{81.37}     & \textbf{89.73}    & \textbf{95.06}    & \textbf{97.13}    & 88.64             \\ \hline
\end{tabular}
}
\end{table}

% \usepackage{graphicx}
\begin{table}[hbt]
\caption{Comparative experimental results on the Kvasir-SEG and ClinicDB datasets(Bold indicates the best)}
\label{tab:kc}
\resizebox{\columnwidth}{!}{%
\begin{tabular}{llllll}
\hline
\textbf{Model} & \textbf{IOU} & \textbf{DSC} & \textbf{Acc} & \textbf{Spe} & \textbf{Sen} \\ \hline
UNetV2~\cite{gao2022multi} & 84 & 91.3 & 97.47 & 99.08 & 88.39 \\
VMUnet~\cite{ruan2024vm} & 80.32 & 89.09 & 96.8 & 98.49 & 87.21 \\
VMUnetV2~\cite{zhang2024vm} & 84.15 & 91.34 & 97.52 & 99.25 & 87.71 \\
\textbf{HMT-Unet} & \textbf{85.66} & \textbf{92.28} & \textbf{97.69} & 98.7 & \textbf{91.95} \\ \hline
UNetV2~\cite{gao2022multi} & 83.85 & 91.21 & 98.59 & 99.16 & 91.99 \\
VMUnet~\cite{ruan2024vm} & 81.95 & 90.08 & 98.42 & 99.18 & 89.73 \\
VMUnetV2~\cite{zhang2024vm} & 89.31 & 94.35 & 99.09 & 99.38 & \textbf{95.64} \\
\textbf{HMT-Unet} & \textbf{90.96} & \textbf{95.26} & \textbf{99.25} & \textbf{99.61} & \textbf{95.05} \\ \hline
\end{tabular}%
}
\end{table}

% \usepackage{graphicx}
\begin{table}[hbt]
\caption{Comparative experimental results on the ColonDB, ETIS and Endoscenedatasets(Bold indicates the best)}
\label{tab:CEE}
\resizebox{\columnwidth}{!}{%
\begin{tabular}{llllll}
\hline
\textbf{Model} & \textbf{IOU} & \textbf{DSC} & \textbf{Acc} & \textbf{Spe} & \textbf{Sen} \\ \hline
UNetV2~\cite{gao2022multi} & 57.29 & 72.85 & 96.19 & 98.43 & 68.46 \\
VMUnet~\cite{ruan2024vm} & 55.28 & 71.2 & 96.02 & 98.45 & 65.89 \\
VMUnetV2~\cite{zhang2024vm} & \textbf{60.98} & \textbf{75.76} & \textbf{96.54} & \textbf{98.46} & 72.68 \\
\textbf{HMT-Unet} & 60.44 & 75.34 & 96.31 & 97.97 & \textbf{75.68} \\ \hline
UNetV2~\cite{gao2022multi} & 71.9 & 83.65 & 98.35 & 98.61 & \textbf{92.96} \\
VMUnet~\cite{ruan2024vm} & 66.41 & 79.81 & 98.26 & 99.33 & 75.79 \\
VMUnetV2~\cite{zhang2024vm} & 70.84 & 82.23 & 97.43 & 96.91 & 91.8 \\
\textbf{HMT-Unet} & \textbf{71.26} & \textbf{83.22} & \textbf{98.4} & \textbf{98.91} & 87.59 \\ \hline
UNetV2~\cite{gao2022multi} & \textbf{82.86} & \textbf{90.63} & 99.34 & 99.54 & 93.82 \\
VMUnet~\cite{ruan2024vm} & 79.55 & 88.61 & 99.2 & 99.44 & 92.46 \\
VMUnetV2~\cite{zhang2024vm} & 80.32 & 89.09 & 99.25 & 99.55 & 90.77 \\
\textbf{HMT-Unet} & 82.01 & 90.12 & 99.29 & 99.41 & \textbf{95.9} \\ \hline
\end{tabular}%
}
\end{table}

% datasource : /root/workspace/code/mamba_all/VM-UNet/results/valid/hmt_unet_isic_all_Sunday_18_August_2024_11_10_00/log/train.info.log
% \begin{table}[ht]
% \centering
% % \setlength{\tabcolsep}{4pt}
% \caption{Comparative experimental results on the ZD-LCI-GIM dataset(Bold indicates the best)}
% \label{tab:contrast}
% \begin{tabular}{llllll}
% \hline
% \textbf{Model}               & \textbf{IOU}           & \textbf{DSC}            & \textbf{Acc}             & \textbf{Spe}   & \textbf{Sen}   \\ \hline
% DeeplabV3~\cite{chen2018encoder}          & 51.71          & 68.17          & 83.71          & 86.77 & 73.81 \\
% Unet~\cite{ronneberger2015u}               & 54.64          & 70.67          & 85.61          & 89.41 & 73.33 \\
% Unet++~\cite{zhou2018unet++}             & 55.35          & 71.26          & 85.57          & 88.64 & 75.65 \\
% SwinUnet~\cite{cao2022swin}           & 54.61          & 70.64          & 84.46          & 86.13 & 79.08 \\
% VM-Unet~\cite{ruan2024vm}            & 56.76          & 72.42          & 85.87          & 88.16 & 78.48 \\
% Unetv2~\cite{peng2023u}             & 53.82          & 69.98          & 84.63          & 87.36 & 75.81 \\
% \textbf{VM-Unetv2} & \textbf{57.02} & \textbf{72.63} & \textbf{86.05} & 88.46 & 78.29 \\ \hline
% \end{tabular}
% \end{table}

% \usepackage{graphicx}
\begin{table}[]
\caption{Comparative experimental results on the ZD-LCI-GIM dataset(Bold indicates the best)}
\label{tab:gim}
\resizebox{\columnwidth}{!}{%
\begin{tabular}{llllll}
\hline
\textbf{Model} & \textbf{IOU} & \textbf{DSC} & \textbf{Acc} & \textbf{Spe} & \textbf{Sen} \\ \hline
DeeplabV3 & 51.71 & 68.17 & 83.71 & 86.77 & 73.81 \\
Unet & 54.64 & 70.67 & 85.61 & 89.41 & 73.33 \\
Unet++ & 55.35 & 71.26 & 85.57 & 88.64 & 75.65 \\
SwinUnet & 54.61 & 70.64 & 84.46 & 86.13 & 79.08 \\
VM-Unet & 56.76 & 72.42 & 85.87 & 88.16 & \textbf{78.48} \\
Unetv2 & 53.82 & 69.98 & 84.63 & 87.36 & 75.81 \\
VM-Unetv2 & 57.02 & 72.63 & 86.05 & 88.46 & 78.29 \\
\textbf{HMT-Unet} & \textbf{57.3} & \textbf{72.86} & \textbf{86.61} & \textbf{89.87} & 76.05 \\ \hline
\end{tabular}%
}
\end{table}

\subsection{Ablation Study}

% \usepackage{multirow}
% \usepackage{graphicx}
\begin{table}[]
\caption{Ablation studies on Init. Weight of MambaVision}
\label{tab:ab}
\resizebox{\columnwidth}{!}{%
\begin{tabular}{c|cc|cc}
\hline
\multirow{2}{*}{Init. Weight} & \multicolumn{2}{c|}{ISIC17} & \multicolumn{2}{c}{ISIC18} \\ \cline{2-5} 
 & IOU & DSC & IOU & DSC \\ \hline
- & 69.63 & 82.1 & 73.13 & 84.6 \\
MambaVision-T & 78.85 & 88.17 & 79.04 & 88.29 \\
MambaVision-S & \textbf{83.05} & \textbf{90.74} & \textbf{81.37} & \textbf{89.73} \\ \hline
\end{tabular}%
}
\end{table}

In this section, we perform ablation studies on the initialization of HMT-UNet, utilizing the ISIC17 and ISIC18 datasets. 
We initiate HMT-UNet by employing no pre-trained weights and pre-trained weights obtained from MambaVision-T and MambaVision-S. The findings from our experiments, as depicted in Table \ref{tab:ab}, demonstrate that the use of more robust pre-trained weights considerably boosts the performance of HMT-UNet in subsequent tasks. This underscores the significant impact that pre-trained weights have on the effectiveness of HMT-UNet.

\section{Conclusion}
\label{conclusion}
In this paper, we introduce a pure MambaVision-based model for medical image segmentation for the first time,
To leverage the capabilities of MambaVision-based models, we construct HMT-Unet using MambaVision Mixer layer blocks and initialize its weights with the pretrained MambaVision-S. Comprehensive experiments are conducted on skin lesion and gastroenterology polyp and GIM segmentation datasets indicate that pure MambaVision-based models are highly competitive in medical image segmentation tasks and merit in-depth exploration in the future.

\bibliographystyle{IEEEtran}
\bibliography{refs}

\end{document}